# Quantification of Cyclic Topology in Polymer Networks Using 3D Nets

Devosmita Sen[1], Bradley D. Olsen[1,*]

[1]Department of Chemical Engineering, Massachusetts Institute of Technology, Cambridge Massachusetts 02139, United States

*Please address correspondence to: bdolsen@mit.edu

**Abstract**

Polymer networks invariably possess topological inhomogeneities in the form of loops and dangling ends. The macroscopic properties of such materials are directly dependent on the local cyclic topology around nodes and chains. Here, a new formalism to model polymer network topology is presented based on the concept of 3D nets that enables computation of these local cyclic topologies. A cycle counting algorithm is developed which characterizes the relevant local topological cycles around every node, enabling comparison of networks to ideal nets based on local cycle distributions. Comparison of networks formed by different simulation algorithms, using a topology-based distance metric, reveals that polymer networks possess a fundamental cycle size which depends on the topological proximity of crosslinkers during bond formation. This parameter identifies distinct topological classes of networks and provides an effective method to simulate a wide array of networks starting from suitable 3D nets. This approach can be readily generalized beyond polymer networks, enabling quantification of the cyclic topology and facilitating study of topology-property correlations.

Polymer networks, from common rubber tires to contact lenses and specialized drug delivery systems [1-10], consist of complex interconnections among constituent polymer chains. Quantitative prediction of macroscopic network properties relies on a fundamental understanding of network topology, which continues to be a challenge. This is because networks inevitably possess topological inhomogeneities in the form of molecular loops [Fig. 1(a)] arising due to intramolecular reactions. Beyond polymer networks, cyclic structures are important for information propagation in internet, social and biological networks [11-13]. Most fundamental network theories are based on the acyclic tree model [Fig. 1(b)] [1,10,14-18]. Cyclic structures have been modelled as independent and isolated loop-defects in a tree-like model [19,20], and Kinetic Graph Theory (KGT) and Kinetic Monte Carlo (KMC) simulations have been successful in describing the density of lower-order loops [21]. It has been shown that the presence of such cycles, especially near the crack tip, critically impacts network properties including elastic modulus, fracture toughness and swelling [19,20,22-31]. Recently, the loop-opening model was proposed [32], which suggests that fracture toughness is directly dependent on the length of the locally shortest loop associated with the bridging chain undergoing scission [Fig. 1(c)]. These studies indicate that the local topology of higher order cycle lengths in networks is critical to determining macroscopic fracture properties. Previous studies have developed algorithms for counting shortest cycles in networks, based on shortest rings and cycle rank considerations, and have characterized global cycle length distributions in simulated networks [33-37]. Such analysis has shown that there must exists a peak in the global cycle length distribution in networks at some intermediate to higher loop order, owing to the constraints of complete conversion and physical space filling constraints in three dimensions [38,39], in contrary to the ideal tree-like assumption. As the size of a tree-network grows, it no longer fits into 3D space and must loop back on itself.

While such a global cycle length distribution gives an estimate of the overall number of cycles in the network, it does not contain complete information about local connectivity patterns between cycles. The complexity of polymer networks suggests that each chain must be part of multiple loops, which leads to cooperative effects between topologically adjacent cycles, affecting the macroscopic fracture behavior. In order to correctly account for such local distribution and connectivity of cycles and accurately predict toughness, a fundamental model for network topology, beyond the tree-like assumption is extremely essential.

This work utilizes a class of ideal structures called three-dimensional nets to model network topology. 3D nets are topologically regular networks formed by interconnecting nodes with edges in a specific arrangement [40-43]. While 3D nets are usually visualized as periodic lattice structures [Fig. 1(d)], its definition does not impose any constraint on the spatial ordering of the node positions, and these structures are solely characterized by their topology. 3D nets consist of overlapping cyclic structures and possess a set of fundamental cycle sizes characterizing their topology. Such network structures can be viewed as idealized models for polymer networks, which inherently lack spatial order and have been shown to contain higher-ordered cycles. Polymer network topology thus corresponds to an intermediate in the spectrum consisting of tree-structures and 3D nets on the two extremes. In this work, a generalized method to characterize network topology has been developed based on 3D nets, providing a systematic approach to quantify higher-order molecular loops and associated local topology distributions.

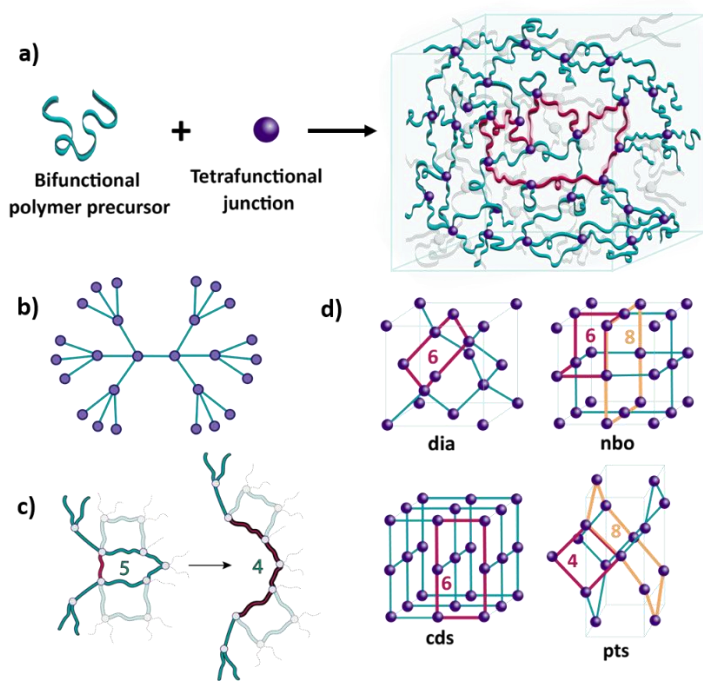

**Figure 1: (a)** End-linking reactions lead to emergence of a network containing higher-ordered cycles **(b)** Infinite-dimension tree structure traditionally used to model networks **(c)** A schematic diagram of how the contour length of opened loop upon scission of a bridging chain, is directly related to the size of the shortest loop around it. Here, the scission of a bridging chain containing a loop of size 5 leads to an opened loop with contour length of 4 **(d)** Commonly found 3D-nets (dia, nbo, cds, pts) named after lattice structures exhibiting these nets, with characteristic fundamental cycle sizes as indicated.

Cyclic topology in 3D nets is characterized by vertex symbols, which quantify the shortest rings around each node [44,45]. A ring is defined as a fundamental cycle that is not the sum of any two shorter cycles [46]. This definition helps to quantify all the relevant shortest cycles around nodes while effectively eliminating redundant cycles. In this work, the concept of vertex symbols in topologically regular 3D net structures has been generalized for random networks. The vertex-symbol of a node with connectivity $p$ is derived by counting all possible ways to unite each of the $\frac{p(p-1)}{2}$ edge-pairs around it using smallest rings, which are added as a subscript index to the ring size. The notation is represented as $M1_{y_1}.M2_{y_2}.M3_{y_3}...$, where $M1, M2, M3..$ are the ring sizes, and $y_1, y_2, y_3, ...$ are the number of rings of that size around the corresponding edge-pair. Rings of size greater or equal to three are counted using vertex symbols ($M1, M2, M3... \geq 3$), and cycles of size one and two are analyzed separately. Here, nodes and edges correspond to crosslinkers and polymer chains respectively. Fig. 2 (a) and (b) show two representative local topologies around central nodes with $p = 4$. For an edge-pair, if there is no cycle satisfying the ring criterion, it is indicated by '*' (see Supporting Information for details). The global cycle length distribution is then characterized by the ratio of the number of loops to the number of chains as a function of the loop order (LO), defined as the number of chains required to close the loop. It is derived by calculating the number of cycles of size $i$ ($\geq 3$) in the network ($N_i$):

$$N_i = nint(\sum_{node\ j} \frac{n_{i,j}(f_j - 2)}{2m_j}) \qquad \ldots\ldots\ldots (1)$$

Where $nint(x)$ is the nearest integer function, and

$$n_{i,j} = \sum_{\substack{element\ k\ of\ vertex \\ symbol\ of\ node\ j\ where\ Mk=i}} y_k|_{Mk=i} \qquad \ldots\ldots\ldots (2)$$

$$m_j = \sum_{element\ k\ of\ vertex\ of\ node\ j} y_k \qquad \ldots\ldots\ldots (3)$$

$f_j$ is the number of chains connected to node $j$. Here, $n_{i,j}$ represents the number of cycles of size $i$ enclosed by edge-pairs of node $j$, and $m_j$ is the total number of cycles around node $j$. For $i = 1$ and 2, the cycles are calculated separately around every node or node-pair respectively.

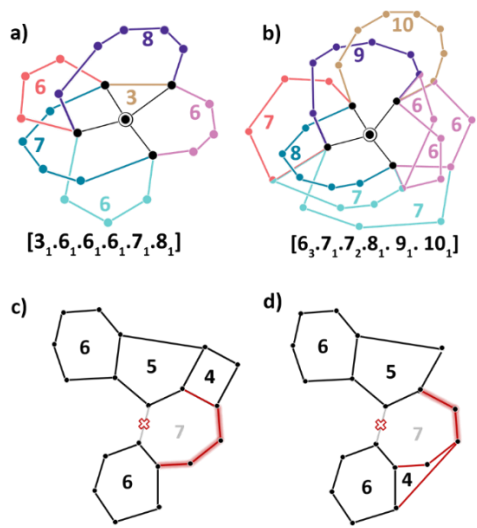

**Figure 2: (a),(b)** Representative local topologies, and their corresponding vertex-symbols. Cycle sizes around each edge-pair are shown in the corresponding color. **(c), (d)** Show two networks having the same global cycle length distribution, but different local distributions. If the chains indicated by crosses in the two networks are broken, the chains along the opened loop have different arrangement of loops around themselves, giving rise to different macroscopic fracture properties. Such differences in local topology can be captured using vertex symbols.

Vertex symbols can thus effectively characterize relevant local topology in networks. Fig. 2(c) and (d) show two networks having the same global cycle length distribution, but different connectivity patterns between cycles, leading to different local topologies. Such a difference is distinctly captured by the vertex symbols around each node. Further, in the event of the scission of the chain indicated, the corresponding chains forming the opened loop have different distribution of cycles around them, leading to different fracture properties of the network. Such connectivity patterns of looped chains among themselves is not readily available from the global cycle length distributions, but can be effectively captured by the local cyclic topology characterized by the vertex symbol.

Here, networks formed via end-linking of bifunctional polymer precursors ($A_2$) and tetrafunctional junctions ($B_4$) are considered [Fig. 1(a)]. Applying this loop counting formalism, it is demonstrated that different generation algorithms produce networks with substantially different underlying topologies. Seven network simulation algorithms were studied as comparative models: KMC [47], Gusev [48], Annable [49], Leung and Eichinger [50], Lei and Liu [51], Grimson [52], and Bond Fluctuation Model [53]. KMC simulations are based on a purely topological perspective, with only average topological distances between crosslinkers being considered. The Gusev, Annable and Leung models are based on real space distributions of crosslinkers and probabilistic chain connections. The Bond Fluctuation Model explicitly accounts for diffusion and collision of reactive groups leading to network formation. The Lei and Grimson models are lattice-based. Details of simulations to generate specific network configurations are included in Supporting Information. For a consistent comparison, all the networks in this work have been simulated at a dimensionless concentration $cR^3$ of 10, where $c$ is the polymer concentration and $R$ denotes the root-mean-square end-to-end distance of chains. Previous work using KGT and KMC, along with experimental evidence has shown that network topology depends on a single dimensionless parameter $cR^3$[21], justifying the use of consistent values across networks. Large enough systems were used to reach asymptotic values of the topological parameters [Fig. S5].

The global cycle-length distribution exhibits a maximum at a certain LO for all networks, similar to observations in previous studies [33-37], indicating an intrinsic LO most commonly observed in each network [Fig. 3(a)]. For the Gusev, Leung, Annable, and BFM models, the distribution peaks at LO = 8. For the KMC network, it peaks at LO = 11. Since the KMC simulation is purely topological and not constrained by density limitations, bonds can form between topologically distant crosslinkers, increasing average cycle sizes to higher orders. For the Lei et al. network, the peak appears at LO = 4. The Grimson model peaks at LO = 6 and contains loops of only even orders, since bond dilution of a cubic lattice only produces even ordered loops that are longer than its initial fundamental cycle size of 4. The latter two networks are formed via nearest-neighbor connections, making the bonds locally aggregated, biasing the LO to lower values.

The effect of local aggregation of bonds is further demonstrated by performing successive chain rearrangements on a topologically regular network, revealing a transition in the peak LO of the global cycle-length distributions. The periodic diamond crystal lattice, a frequently used model for polymer networks [54,55], is used as the starting structure. The crosslinkers are positioned on the lattice sites, and the polymer chains along the nearest-neighbor connections. Topological rearrangements are performed by rewiring bonds between chains and crosslinkers based on the Watts-Strogatz model [56], a systematic method to study lattice to random transitions in networks. Two chains are selected randomly, and one of each of their connecting crosslinkers are interchanged, subject to the condition that the length of the two new chains after exchange is less than a specified length $d$. The polymer precursor A$_2$ is modelled as a monodisperse, flexible Gaussian chain with number of Kuhn segments $N = 12$ and segment length $b = 1$ unit. Hence, the contour length of the chain is $L_C = 12$ units. For $cR^3 = 10$, the lattice parameter is $0.337L_c$, with nearest-neighbor and second nearest-neighbor distances as $0.146L_c$ and $0.239L_c$, respectively (see Supporting Information). The swaps are continued until the global cycle-length distribution is no longer altered by the rearrangements [Fig. 3(b)]. The distribution shows a clear transition from LO equal to the initial fundamental cycle size of 6 of the diamond lattice to higher values. A new peak emerges at LO=11. Performing chain rearrangements with stricter constrains by decreasing the value of $d$ shows a shift in the peak position of LO to lower values [Fig. 3(c)]. Networks were generated with $d= L_C, \frac{L_C}{2}, \frac{L_C}{3}$ and $\frac{L_C}{4}$, referred to as DWS_1, DWS_2, DWS_3 and DWS_4 networks, respectively. For DWS_4, the distribution peaks at LO=7, unlike DWS_1 (peak LO=11). Hence, biasing chain connections to be more localized by tuning the parameter $d$ shifts the peak LO towards shorter cycles. Physically, as $d$ decreases, the connections resulting from the edge-swaps are formed between nodes that are nearer to each other, restricting the topological distance between the nodes on cycles and shifting the LO to smaller values. This analysis is consistent with the trends in Fig. 3(a). In the Lei and Grimson models, chain connections predominantly arose from nearest-neighbor connections, yielding a peak at low LO. On the other hand, the absence of density limitations in the KMC network allowed the chains to connect between topologically-distant crosslinkers, shifting the peak to higher LO.

Increasing crosslinker functionality ($p$) also leads to a preference for shorter loops due to the increased probability of a chain-end connecting back to the same crosslinker as the other end. KMC networks simulated for $p$ values ranging from 3 to 8 [Fig. 3(d)], show a shift from a peak LO=17 in tri-functional networks to LO=7 in octa-functional networks. With increase in $p$, the distribution becomes sharper, and the total number of loops increases. The number of edge-pairs available to form cycles increases; therefore, a single chain can be part of a greater number of cycles, increasing the total number of cycles relative to the number of chains in the network.

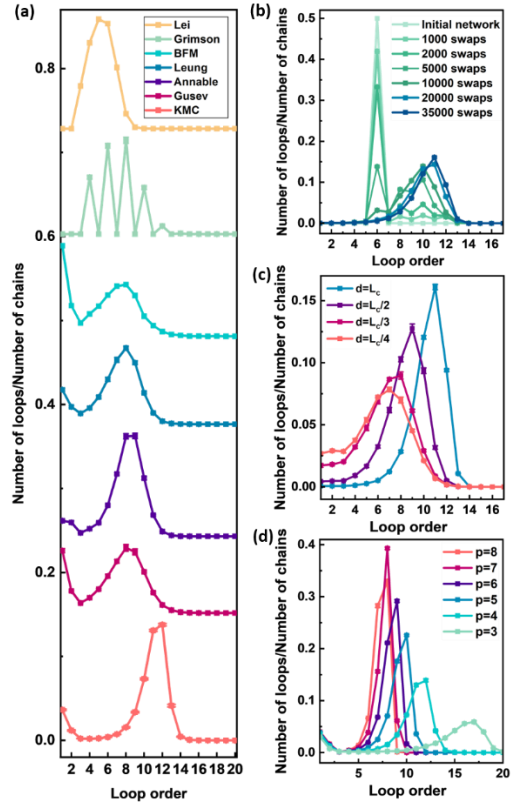

**Figure 3:** Offset plot of global cycle-length distribution, for networks generated by **(a)** KMC, Gusev, Annable, Leung, BFM, Lei and Grimson models, with the offsets of 0, 0.15, 0.24, 0.38, 0.48, 0.60, and 0.73 respectively **(b)** by transition of network from pure diamond topology to a randomized network for $d = L_c$. **(c)** by DWS_1, DWS_2, DWS_3 and DWS_4 networks. **(d)** by KMC simulations for junction functionalities ranging from $p = 3$ to $p = 8$. Error bars represent standard deviation over 10 replicates.

To investigate the effect of localization of chain connections and clustering on the local cyclic topology, the dissimilarity of networks is quantified. The vertex-symbol notation quantifies the local cyclic structure around nodes, enabling a direct node-to-node comparison between networks. Using this, normalized frequency distributions of vertex symbols are obtained for all networks considered in this work. The first few most commonly observed elements of the distribution for a sample of each network are shown in Fig. 4(a),(b) and Fig. S4(a)-(h). For purposes of computational speed, the vertex-symbols are lumped by dropping the subscript indices, and only the ring sizes are used for further analysis. While the observed cycle sizes are different for each network, these distributions show a common trend - even though local cyclic topology is not perfectly regular across every node, the cycle sizes differ from one another by only one or two in most cases, and are very close to a fundamental cycle size, characterized by the corresponding peak LO of the global cycle-length distribution of that network.

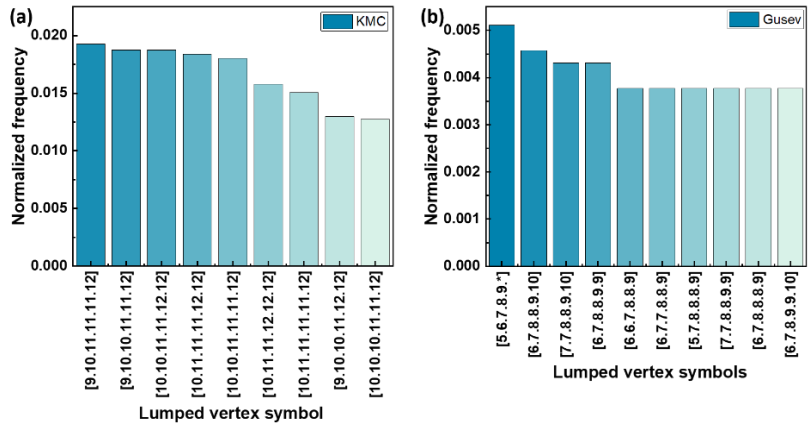

**Figure 4:** First few most common vertex-symbols observed in a sample of **(a)** KMC, **(b)** Gusev network

Such differences in local topology are quantified by utilizing the concept of Earth Mover's Distance (EMD) [57,58] to obtain a distance-like dissimilarity metric between the normalized frequency distributions of vertex symbols. A new metric, the Loop Edit Distance (LED) is developed for quantifying distances in the EMD calculation (Refer to Supporting Information for details). Physically, the EMD between two networks measures the minimum number of loop edits required to transform the local topology around each node of a network to that of the other, representing the overall dissimilarity between the network pair. EMD values, calculated for all binary pairs of trials for each network pair [Fig. 5(a)], show standard deviations less than 5% in all cases, indicating high reproducibility. Performing a hierarchical clustering analysis on the pairwise EMD (Supporting Information Section X) reveals three different network categories- (i) KMC and DWS_1; (ii) Annable, Gusev, Leung, and DWS_2; and (iii) BFM, Grimson, Lei, DWS_3 and DWS_4 networks. These categories correspond to the networks having the lowest to the highest extent of bond clustering, respectively, affecting the LO vis-à-vis the overall topology. Such a classification suggests clear distinctions between the local cyclic topologies observed in networks generated by different simulation algorithms, suggesting distinct macroscopic fracture properties of these networks.

The deviation of the DWS network topologies from a regular diamond lattice, as quantified by the EMD values, decreases with an increase in the extent of chain clustering [Fig. 5(b)]. The EMD is highest for DWS_1, where chain connections between topologically distant crosslinkers up to the contour length are allowed, and decreases as the topological proximity of crosslinkers across chain connections increases. It eventually saturates and increases slightly for DWS_4, owing to cycles of lower orders being preferred, which slightly increases the dissimilarity even though the peak LO gets closer to the fundamental cycle size of the original diamond topology.

Similar to the diamond lattice network considered here, this analysis can be applied to any 3D net with desired topology, and chain rearrangements can be used to control topological regularity, quantified by the EMD values. This opens up opportunities for simulating a wide array of polymer networks with desired topological features starting from any 3D net.

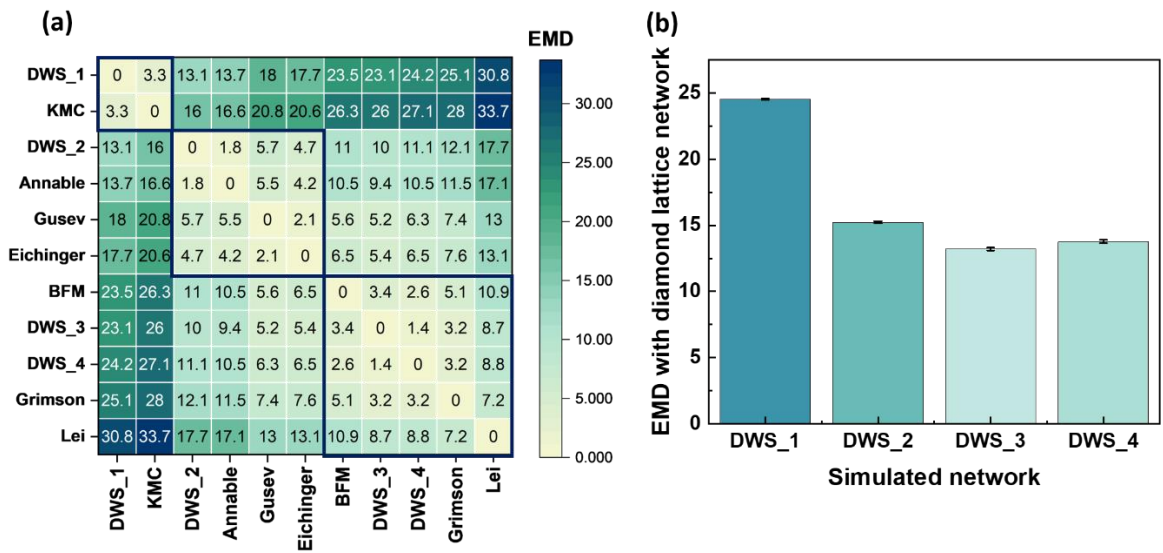

**Figure 5: (a)** Heatmap of EMD comparison between networks considered in this work. The 3 network classes are shown by bolded boxes. **(b)** EMD comparison of DWS_1, DWS_2, DWS_3 and DWS_4 networks with pure diamond lattice, with standard deviations less than 1% in all cases

In summary, a new model for polymer network topology is proposed in this work based on the concept of 3D nets. A cycle counting algorithm, capable of exactly quantifying local topology around nodes, is developed based on vertex-symbol notations of 3D nets. Deriving global cycle length distributions reveals that the peak LO for networks is dependent on the inherent physics of the simulation algorithm. Comparing EMD values between network topologies using the LED distance metric elucidates that cyclic topology is governed by the topological proximity of crosslinkers across chains during bond formation. The constraint of chain aggregation favors shorter loops; when constraints are relaxed, the loop size increases, achieving the largest values when spatial packing constraints are ignored. This formalism provides a systematic approach to quantify local cyclic topology in networks that can be directly correlated to macroscopic properties. By tuning the topological proximity between crosslinkers, a wide array of networks with diverse topological characteristics can be simulated starting from any 3D net structure, and can be suitably generalized to more complex interconnected systems beyond polymer networks.

**Acknowledgement**

This work is supported by National Science Foundation (CHE-2203951). The authors acknowledge the MIT SuperCloud and Lincoln Laboratory Supercomputing Center for providing high performance computing resources. We thank Jiale Shi and Brilant Kasami for discussions and suggestions.

# Supporting Information

**Quantification of Cyclic Topology in Polymer Networks using 3D Nets**

Devosmita Sen[1], Bradley D. Olsen[1,*]

[1]Department of Chemical Engineering, Massachusetts Institute of Technology, Cambridge Massachusetts 02139, United States

*Please address correspondence to: bdolsen@mit.edu

## I. Examples of calculation of vertex symbols for some representative topologies

Fig. S1 (a), (b) show two representative topologies and calculation of their corresponding vertex symbols. In Fig S1 (a), edge-pair (1-5,1-2) constitutes a cycle of 6 shown in red, and edge-pair (1-5,1-3) constitutes a cycle of 8, shown in violet. Similarly, the remaining edge-pairs around node 1, and their corresponding cycles are (1-2,1-3):3, (1-3,1-4):6, (1-4,1-5):6, (1-5,1-2):6, (1-2,1-4):7. In Fig. S1 (b), both edge-pairs (1-2,1-5) and (1-2,1-3) constitute rings of size 6. Hence, even though the edge-pair (1-3,1-5) contains an 8-membered cycle, it is the sum of the two 6-membered cycles already counted and is thus redundant. Therefore, the vertex-symbol is assigned as $[3_1.6_1.6_1.6_1.7_1.*]$.

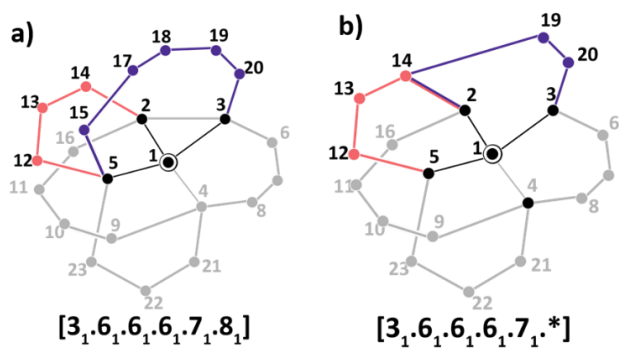

$[3_1.6_1.6_1.6_1.7_1.8_1]$   $[3_1.6_1.6_1.6_1.7_1.*]$

**Figure S1: (a)** Demonstration of loop counting in networks using vertex symbols. **(b)** Modification of the network in (a) demonstrating the use of '*' symbol.

## II. Derivation of global cycle length distribution from vertex symbols

The number of cycles of size $i$ in a network ($N_i$), is derived from the vertex symbols on each node (equation 1,2,3 in main text). Hence, the total number of cycles in the network can be obtained as:

$$N_{tot} = \sum_i nint(\sum_{node\ j} \frac{n_{i,j}(f_j - 2)}{2m_j})$$

$$\approx \sum_{node\ j} \left(\frac{(f_j - 2)}{2m_j}\right) \sum_i n_{i,j} = \sum_{node\ j} \left(\frac{(f_j - 2)}{2m_j}\right) m_j \approx \frac{v(f - 2)}{2}$$

Where $v$ is the number of nodes (vertices) in the network, and $f$ is the overall functionality of crosslinks in the network.

The cycle rank in graph theory is defined as:

$$N_{tot,cr} = e - v + 1$$

Where, $e$ = number of edges (chains), $v$ = number of vertices (nodes)

For a fully connected network containing bifunctional chains, $e = \frac{f}{2}v$

Hence, the cycle rank is: $N_{tot,cr} = \frac{v(f-2)}{2} + 1$

For large system sizes where $N_{tot,cr} \gg 1$, $N_{tot,cr} \approx N_{tot}$

The total number of cycles obtained from this algorithm undercounts the cycle rank by approximately 1 cycle. Since $N_{tot}$ scales the same way as the cycle rank, this small underestimation is insignificant in the limit of large networks, making the global cycle length distribution from this analysis consistent with the cycle rank from graph theory. The total number of cycles in the large networks simulated using different models considered in this work has been calculated to be within 0.05% error tolerance of the cycle rank.

Comparison to algorithm by Lang et al:

The comparison of the global cycle length distribution from our algorithm with the algorithm proposed in *M. Lang, W. Michalke, S. Kreitmeier, Macromol. Theory Simul. 10(2001) 204-208*, for the KMC network with N=10000 as shown in Fig S2 (a).

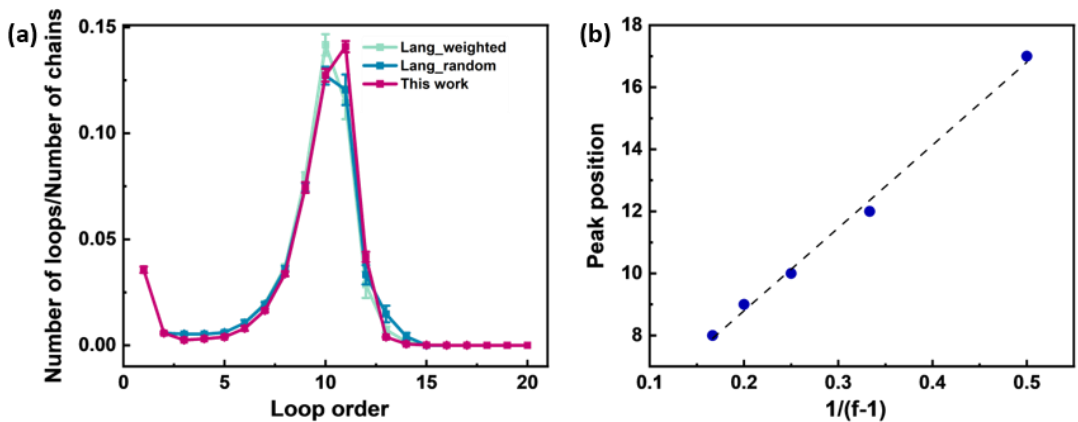

**Figure S2: (a)** Comparison of global cycle length distributions obtained from the current algorithm to that obtained by previously proposed algorithm by Lang et al. **(b)** Plot of peak position of the global cycle length distribution as a function of $\frac{1}{f-1}$ shows a linear dependency, consistent with the analytical scaling relation proposed by previous works. The black dotted line shows the linear fit.

Here, 'Lang_random' refers to the case where random weighting of edges are utilized for constructing the spanning tree in the loop counting algorithm. 'Lang_weighted' refers to weighting of edges according to the criterion of length of shortest mesh an edge is part of. The global cycle size distributions are very similar, and this analysis shows that the loop counting algorithm proposed in this work is consistent with approaches proposed earlier in the field. Since previously-proposed algorithms involve heuristic measures for calculation of loop distributions, the peak position can vary by about 1 based on these metrics, and our algorithm is consistent within the limit of variation of such previously proposed algorithms. The algorithm proposed in our current work poses an advantage that the global cycle length distribution is not dependent on any heuristic measure, and hence, always provides a definitive, absolute loop size distribution of the simulated networks.

Further, the position of the peak in the global cycle length distribution has been compared with the theory proposed in *M. Lang, W. Michalke, S. Kreitmeier,* J. Chem. Phys. **114** *(2001) 7627-7632* and *M. Lang, S. Kreitmeier, and D. Göritz, Rubber Chemistry and Technology **80**, 873 (2007).* These works have shown that the peak loop size of the distribution is approximately proportional to $\frac{1}{f-1}$, where $f$ is the junction functionality. Comparison of the simulation data from the KMC algorithm with this analytical scaling relation shows good agreement. The plot of the peak position from our method as a function of $\frac{1}{f-1}$ has a linear dependency, as shown in Fig S2(b), which confirms that the current approach is consistent with previous works in the field.

The following table (Table S1) considers a few example networks, and the corresponding global cycle size distributions obtained by the algorithm proposed in this work as well as Lang's algorithm. The total number of cycles obtained using the algorithm proposed in this work undercounts the cycle rank by at most 1, which is statistically insignnificant for large networks. While Lang's algorithm always correctly estimates the total numbe of cycles in the network, the actual loop size distribution in the network is dependent on the heuritstics, as clearly seen from Fig S2(a) and row 4 (KMC network) of Table S1. The difference arises due to the overlapping of cycles on one another, thus biasing the cycle sizes based on the underlying heuristics. Further, the cycle size distribution estimated for the *nbo* net using Lang's algorithm (row 3 Table S1) differs from that estimated using net theory, thus highlighting the dependence on the inherent heuristics of the algorithm.

**Table S1:** Comparison of cycle size distribution with Lang's algorithm for a few example networks

| **Example network** | **Cycle size distribution from the algorithm proposed in this work** | **Cycle size distribution from Lang's algorithm** | **Cycle rank** |
|---|---|---|---|
| (network diagram) | (plot) Total number of calculated cycles = 3 | (plot) Total number of calculated cycles = 3 | 3 |

| | | | 4 |
|---|---|---|---|
| 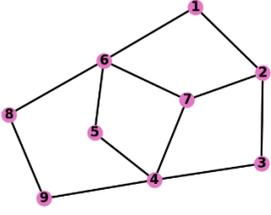 | 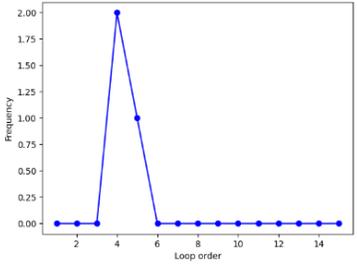  Total number of calculated cycles = 3 | 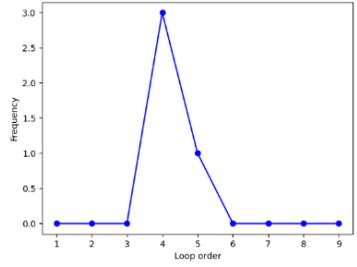  Total number of calculated cycles = 4 | |
| Periodic *nbo* net lattice, with number of nodes $n_{nodes} = 750$  Every node has vertex symbol $6_2.6_2.6_2.6_2.8_2.8_2$  Hence, according to net theory, there will be $\frac{2*n_{nodes}}{3}$ loops of order 6, and $\frac{n_{nodes}}{3}$ loops of order 8 | 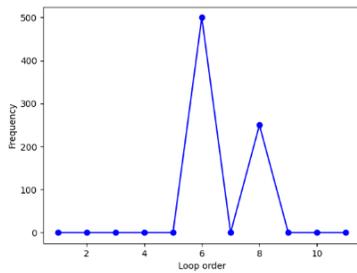  Total number of calculated cycles = 750 | 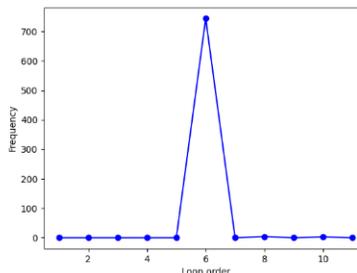  Total number of calculated cycles = 751 | 751 |

| KMC network, with number of chains $n_{chains} = 250$ | 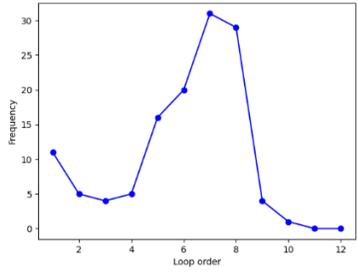 Total number of calculated cycles = 126 | 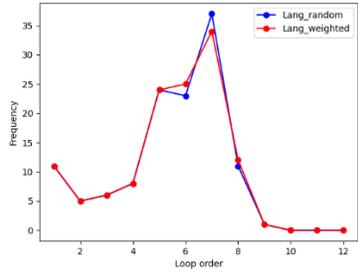 Total number of calculated cycles = 126 | 126 |

The examples highlighted in Table S1 show Lang's algorithm for the case where chains are added to the spanning tree using an optimized algorithm based on the shortest mesh size. Fig S3 (a) and (b) show how the cycle lengths calculated in a network vary based on the extent of optimization of the method of adding chains, even though the cycle rank is the same in both cases. The cycle lengths calculated by Lang's algorithm are based on the underlying heiristics used during the construction of the spanning tree and sequenctially adding chains for mesh quantification.

The algorithm proposed in this work provides the advantage that the global cycle size distribution is deterministic for a given network, and is independent of any heuristic.

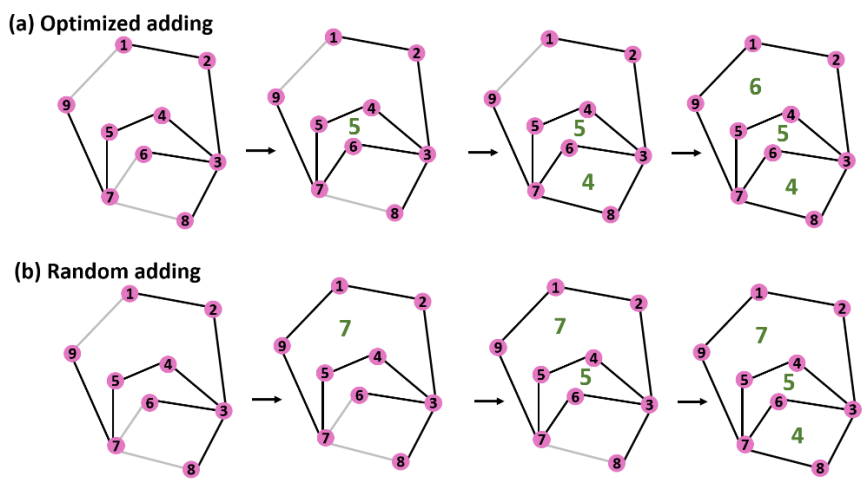

**Figure S3**: The global cycle length distribution obtained using Lang's algorithm based on two different heuristics of adding chains. Black lines indicate the edges that are part of the initial spanning tree and the network as the chains are added to close meshes- (a) Optimized adding- where chains are added back to the network using an optimized method based on the shortest mesh size- give cycles of size 4,5 and 6, and (b) Random adding- where chains are randomly added back to the network- give cycles of size 4, 5 and 7. The cycle lengths calculated in the two cases are different, showing the dependence of the global cycle length distribution on the inherent heuristics used in the algorithm.

## III. Frequency distribution of lumped vertex symbols:

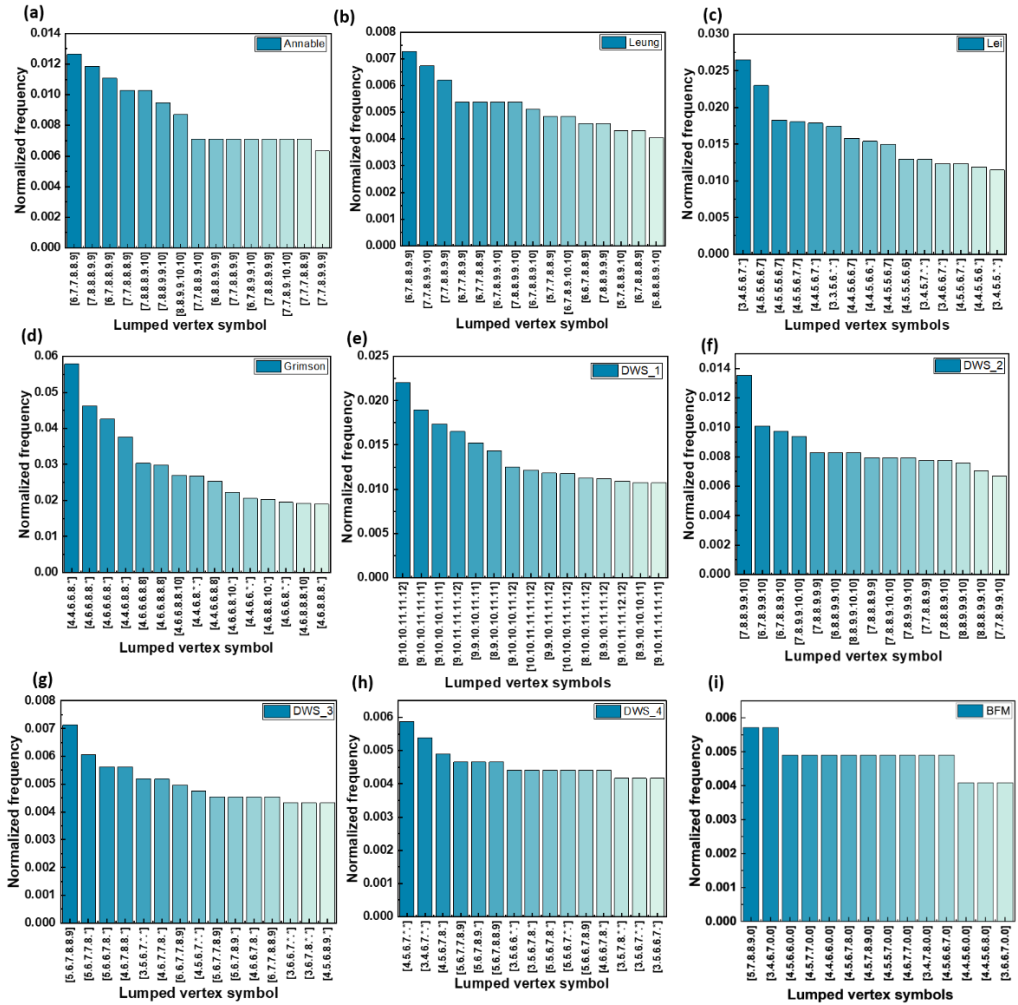

**Figure S4**: First few elements of the normalized frequency distribution for the following networks: **(a)** Annable, **(b)** Leung, **(c)** Lei, **(d)** Grimson, **(e)** DWS_1, **(f)** DWS_2, **(g)** DWS_3, **(h)** DWS_4, **(i)** BFM

The number of vertex-symbols observed in each simulated network is large and widely distributed, demonstrating that, as expected, polymer networks do not exhibit strong topological regularity. In order to coarse-grain the system for purposes of computational speed, the vertex-symbols are lumped by dropping the subscript indices, and only the ring sizes are used for further analysis. Therefore, the lumped vertex-symbol is of the form $A.B.C.D.E.F$, where letters $A$ to $F$ denote the ring sizes in ascending order. In case of a $'*'$ symbol, the corresponding ring size is assigned as 0 and added at the end of the vertex-symbol.

These distributions show that that the KMC and DWS_1 networks have most ring sizes in the range 9-12; the Gusev, Annable, Leung and DWS_2 networks in the range 7-10; and the Lei, Grimson, DWS_3 and DWS_4 in the range 4-8.

IV. **Calculation of Earth Mover's Distance (EMD):**

The EMD is calculated by finding a flow $F = [f_{i,j}]$ which minimizes a cost function.

$$EMD = \frac{\min_F \sum_{i=1}^{m} \sum_{j=1}^{n} f_{i,j} d_{i,j}}{\sum_{i=1}^{m} \sum_{j=1}^{n} f_{i,j}} \quad \ldots\ldots\ldots (S1)$$

Here, cluster representatives $i$ and $j$ correspond to the local cyclic topology around nodes given by the lumped vertex-symbols. $F = [f_{i,j}]$ represents the optimal flow, ie., the amount of weight at $i$ that has to be transported to $j$ in order to transform one distribution to the other. $D = [d_{i,j}]$ represents the distance between $i$ and $j$. Here, a new distance metric, the Loop Edit Distance (LED), is developed for estimating $d_{i,j}$. LED is defined as the minimum number of edits, equivalently, the net number of edge deletions and insertions, required to transform the local cyclic topology of node $i$ to that of node $j$. Hence, a unit change in LO corresponds to one edit. The LED between two nodes having lumped vertex-symbols as $A.B.C.D.E.F$ and $G.H.I.J.K.L$ is defined as-

$$LED = |A - G| + |B - H| + |C - I| + |D - J| + |E - K| + |F - L| \quad \ldots\ldots\ldots (4)$$

Distinct network classes are identified by comparing the topological similarity of networks using the lumped vertex-symbol distributions.

## V. System size scaling:

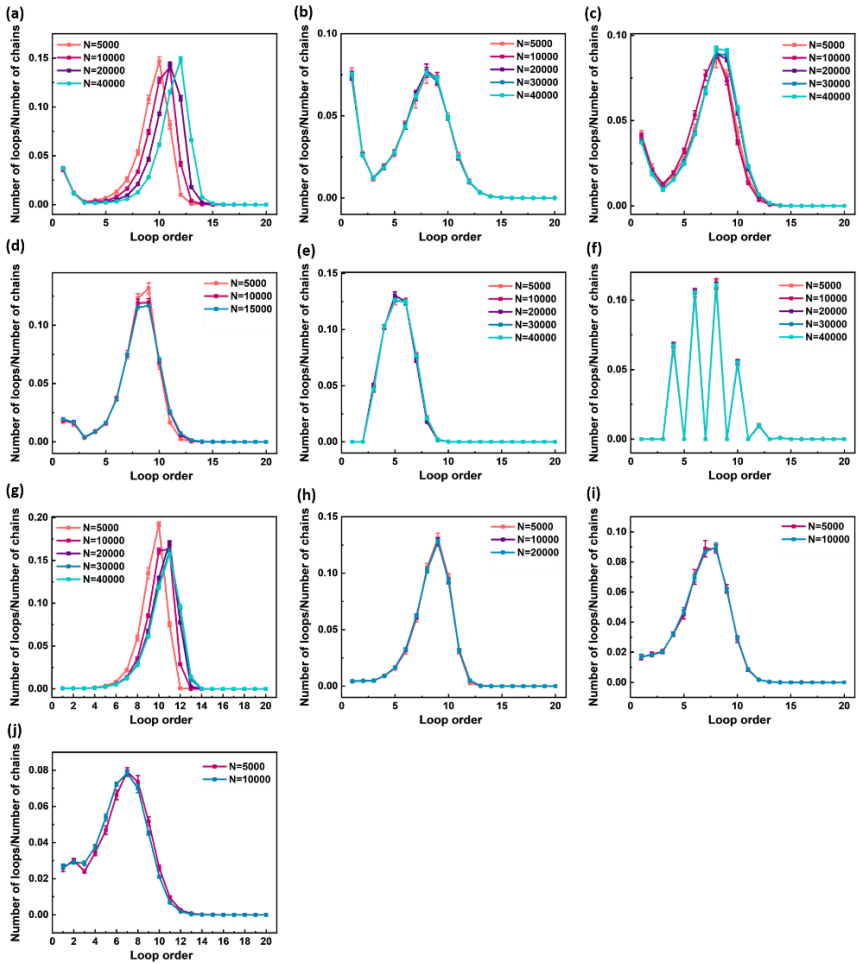

**Figure S5**: Global cyclic topology distribution as a function of the system size scaling for all the networks considered in this work. **(a)** KMC, **(b)** Gusev, **(c)** Leung, **(d)** Annable, **(e)** Lei, **(f)** Grimson, **(g)** DWS_1, **(h)** DWS_2, **(i)** DWS_3, **(j)** DWS_4

## VI. Details of Simulation Algorithms:

1. <u>Kinetic Monte Carlo</u> [1]: This method is based on a purely topological perspective, where the spatial positions of the polymers and junctions are ignored. Only topological distances between the reactive groups, *i.e.*, the number of polymer chains in the shortest topological pathway connecting two groups are tracked. The bifunctional linear polymer $A_2$ is modeled as a flexible Gaussian chain, with two 'A' functional groups located at the respective ends of the chain. Networks are constructed by sequentially selecting a pair of unreacted chain end (A group) and unreacted crosslinker site (B group) at each step with a probability $P_{AB}$, reflecting the topological connections between these two groups:

$$P_{AB} = \frac{\frac{1}{V} + \left(\frac{3}{2\pi R^2 d_{AB}}\right)^{3/2}}{\sum_{ij}\left[\frac{1}{V} + \left(\frac{3}{2\pi R^2 d_{ij}}\right)^{3/2}\right]}$$

where the summations are over all unreacted A-B pairs, $d_{ij}$ is the topological distance connecting the $i^{th}$ A group and the $j^{th}$ B group, $R$ is the root-mean-square end-to-end distance of the Gaussian $A_2$ polymer chains chain, and $V$ is the system volume calculated according to the set concentration $c$. Since this algorithm is based on a purely topological perspective, it is expected that the loop order distribution will never fully converge with respect to system size, and the peak will shift to higher LO with system size. In this work, $N_A = 30000$ has been chosen [Fig S5(a)].

2. <u>Gusev</u> [2]: Positions of crosslinkers in the simulation box are initially assigned according to an ideal gas distribution. Each network chain is added sequentially by reacting one of its ends with a randomly selected cross-linker and reacting the other end with one of the available cross-linkers, which in turn is selected randomly from the Gaussian distribution of distances of all of the available cross-linkers from the already-reacted end of the strand. Periodic boundary conditions are imposed in all three directions. $N_A = 10000$ has been chosen based on the system size scaling [Fig S5(b)].

Code validation:

For networks generated using the parameters: Number of strands per unit volume $(\nu) = 0.5\ nm^{-1}$, crosslinker functionality $\phi = 3$, and probability of reaction between a chain-end and a crosslinker $\xi = 0.9$, the number of free + dangling ends obtained from the current implementation of the code is equal to $0.189 \pm 0.003$, consistent with value of $0.19$ as reported by Gusev[2].

3. <u>Annable</u> [3]: This simulation algorithm was originally proposed for modelling dynamic associative networks, and it has been adapted for tetrafunctional end-linked networks in this work. Here, it simulates the effect of associative dynamics on the network by allowing chain rearrangements. The initial network configuration is generated similar to the Gusev network, and then association steps are performed iteratively. One end of a chain is exchanged with a crosslinker with a *Monte Carlo* algorithm based on chain energies, calculated using the Mao model [4]. The exchanges are continued until the acceptance rate and the mean end-to-end distance converges to a saturation value. $N_A = 10000$ has been chosen based on the system size scaling [Fig S5(d)].

Simulation convergence results:

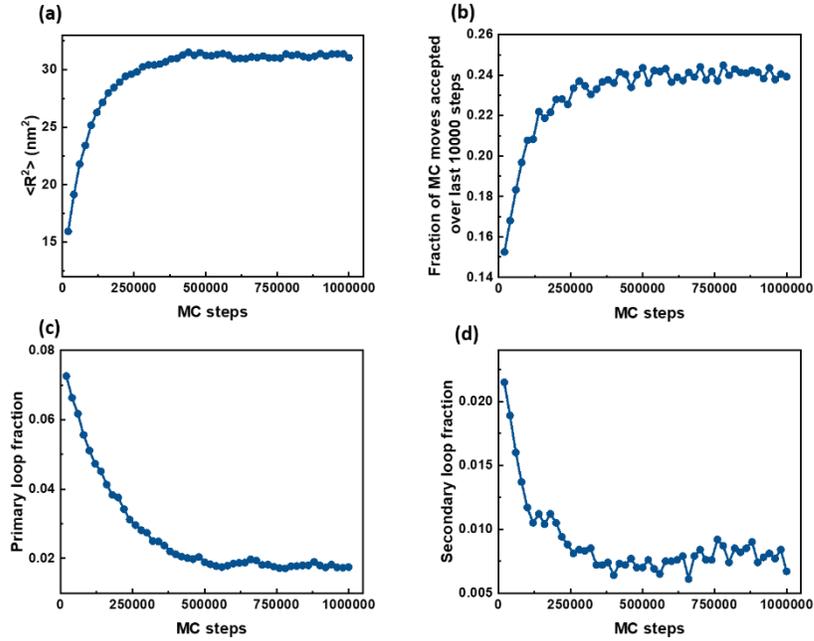

**Figure S6:** Convergence results for the Annable network simulation. **(a)** Mean squared end-to-end distance **(b)** Fraction of MC moves accepted over last 10000 steps over the course of the simulation **(c)** Primary loop fraction **(d)** Secondary loop fraction

4. <u>Leung and Eichinger</u> [5]: The positions of crosslinkers in the simulation box are initially assigned according to an ideal gas distribution. The network is grown by connecting crosslinkers with a chain in increasing order of the distance $r$ between the two crosslinkers, provided $r$ is less than a specified value $d$, which is taken to be the contour length in this work. $N_A = 10000$ has been chosen based on the system size scaling [Fig S5(c)]. The originally proposed algorithm of Leung and Eichinger [5] also showed that the results become stationary for samples with more that 5000 chains, which justifies the use of $N_A = 10000$ in our analysis.

Code validation:

The fraction of chains which constitute inner loop structures in the gel component of the network have been compared for validating the current implementation of the model. Inner loop structures are subgraphs with fully saturated crosslinkers containing exactly one primary loop, and the other two chains connected to different crosslinkers. Inner loop fraction in the gel component as a function of the extent of reaction is shown in Fig S7. Different extents of reactions have been obtained by tuning the value of the parameter $d$.

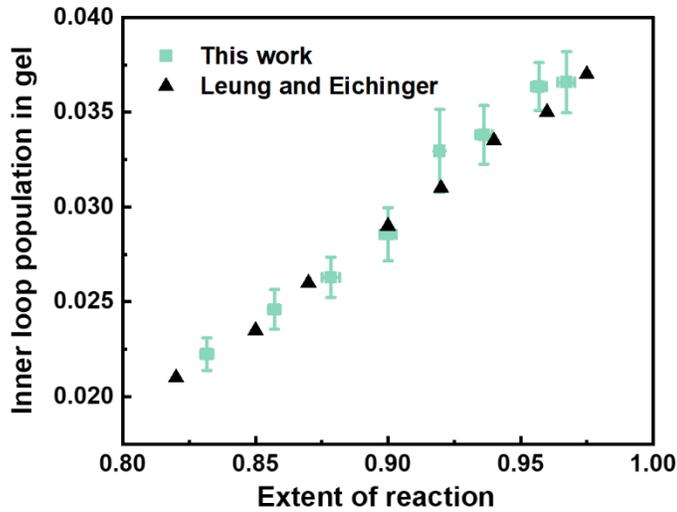

**Figure S7:** Population of inner loop in the gel component of the network as a function of the extent of reaction for $cR^3 = 10$, as obtained in the original work by Leung and Eichinger, and the current implementation of the model in this work

5. Bond Fluctuation Model [6]: This algorithm was first introduced by Carmesin and Kremer [7] in two dimensions and expanded to three dimensions [8]. The open source Lattice-based extensible Monte Carlo Algorithm and Development Environment (LeMonADE) framework [6] was used to generate the network. The system parameters chosen are volume fraction $\phi = 0.5$ and $N_A = 5000$.

6. <u>Lei et al.</u> [9]: The network is geometrically modelled using a tetrahedron mesh, where the body centers of the tetrahedral elements represent the crosslinkers and the connections between the body centers of a pair of coplanar tetrahedron elements represent polymer chains. Initial mesh nodes are randomly sampled in the model domain according to a uniform probability density function. Next, the mesh nodes are used to generate a unique tetrahedron mesh using a 3D Delaunay triangulation method. In the present work, the Python module `scipy.spatial.Delaunay` is used for the mesh generation, and periodic boundary conditions have been implemented in the model to minimize any finite system size effect. The topology of this network is concentration-independent. $N_A = 10000$ has been chosen based on the system size scaling [Fig S5(e)].

7. <u>Grimson</u> [10]: An initial cubic lattice network is considered, and a diluted lattice is generated by randomly removing bonds subject to the constraint that the randomly selected bond is only removed if at least one of the nodes at either end of the bond has a connectivity greater than the desired connectivity $p$, which is equal to 4 in this work. Bond dilution proceeds until all the nodes of the network have a connectivity less than or equal to the maximum value $p$. Periodic boundary conditions have been implemented for the initial lattice network. The topology of this network is concentration-independent. $N_A = 10000$ has been chosen based on the system size scaling [Fig S5(f)].

Code validation:

For a 12 x 12 x 12 simple cubic lattice network, random bond dilution procedure results in a mean bond fraction = 0.589 ± 0.001, consistent with the value reported by Grimson (0.589 ± 0.002).

The comparison of the node connectivity of the resulting network is shown in Fig S8, and validates the current implementation of the model.

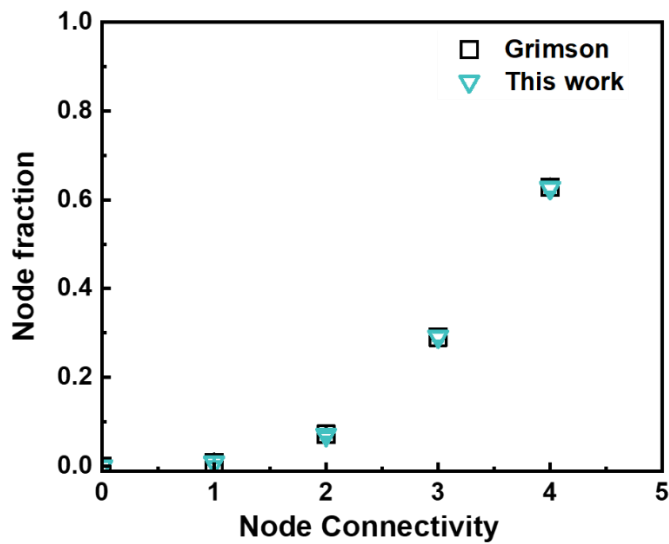

**Figure S8:** Fraction of nodes as a function of node connectivity, as obtained in the original work by Grimson and the current implementation of the model in this work

8. <u>DWS networks</u>: Based on the system size scaling shown in Fig S5(g)-(j), $N_A = 30000$ is chosen for DWS_1 and $N_A = 10000$ is chosen for DWS_2, DWS_3 and DWS_4 networks.

For all simulated networks, since the spatial positions of crosslinkers after network formation do not alter the topology, no force-equilibration-step is necessary for the topology analysis in this work.

## VII. Key approximations in simulation models used in this work:

Different models, with varying levels of approximations, were considered in this work to sample a wide range of networks having different topological properties. These approximations were developed in literature to model specific aspects of the physics of polymer networks while enhancing the computational efficacy of the simulation method. The KMC simulation method is purely topological, and does not have any spatial considerations of crosslinker positions explicitly. This approach is valid for the case where network formation is kinetically limited and diffusion rates are very fast. However, due to this inherent assumption, this model cannot span a wide range of network topologies. Thus, the Gusev, Leung and Annable network algorithms are consdiered, which explicitly incorporate real space position of crosslinkers, and assumed crosslinking distributions. The Annable algorithm models dynamic networks, which represent another potential topological network class. Finally, the Lei and Grimson lattice-based models were considered, which gives a class of networks that originate from a topologically periodic lattice, and are also computationally less expensive, thus providing another set of networks which can potentially have different topological properties. Finally, the Bond Fluctuation Model explicitly considers diffusion and collision, thus providing another alternative approach to model polymer networks.

This is summarized in the table below:

**Table S2:** Approximations of different simulation models considered in this work

| Simulation model | Kinetics | Diffusion | Lattice-based | Excluded volume effects | Extent of coarse-graining | Crosslinking based on spatial position | Pre-assumption about crosslinking distributions |
|---|---|---|---|---|---|---|---|
| Kinetic Monte Carlo (KMC) | Purely topological, valid for kinetically controlled | Diffusion of chains is infinitely fast | No spatial positions of crosslinkers are utilized, network | No excluded volume effects considered since there is | Coarse-grained on the chain-crosslinker level. No information of individual | Crosslinking is purely based on topological distances, no | Crosslinking purely based on topological distances, no prior |
| | network formation regime | | formation is purely based on topology, and hence no lattice considered | no spatial information | monomers is tracked | spatial information utilized | crosslinking distributions assumed |
| Gusev and Leung | Crosslinking is based on the available neighbors around a crosslinker at every stage of crosslinking, thus partially incorporating a kinetic effect. | Diffusion of chains is not explicitly considered | Crosslinker positions in space can vary freely in 3D space and are not based on any underlying lattice | No excluded volume effects of chains and crosslinkers are considered | Coarse-grained on the chain-crosslinker level. No information of individual monomers is tracked | Crosslinking is based on the positions of available neighbors around a crosslinker | Crosslinkers positions are assumed to be distributed uniformly randomly prior to crosslinking. |
| Annable | Initial network formation is similar to Gusev. The next step of bond exchange is dependent on the number of available unreacted crosslinkers in the system, thus incorporating kinetics | Bond exchange step involves rearrangement of chains both in terms of spatial positions and topology, thus partially incorporating the effect of diffusion | Crosslinker positions in space can vary freely in 3D space and are not based on any underlying lattice | No excluded volume effects of chains and crosslinkers are considered | Coarse-grained on the chain-crosslinker level. No information of individual monomers is tracked | Crosslinking is based on the positions of available neighbors around a crosslinker | Initial network formation is similar to Gusev, however, chain rearrangements help to reduce the bias towards the dependence of topology on the pre-crosslinking distribution |
| Lei and Grimson | These models are based on the structure of topologically periodic lattice, thus not incorporating any reaction kinetics | These models are based on the structure of topologically periodic lattice, thus not incorporating any diffusion of chains | These models are based on the structure of topologically periodic tetrahedral and bond diluted cubic lattice respectively, and are thus purely lattice based. | The structure of periodic lattices require that no crosslinker positions can overlap, thus implicitly incorporating excluded volume effects | Coarse-grained on the chain-crosslinker level. No information of individual monomers is tracked | Crosslinking is based on the nearest neighbor connections in lattices, thus crosslinking depends on the spatial positions of nodes in lattices | All crosslinking is dependent on the nearest-neighbor connections in lattices, thus fully biasing the crosslinking distributions |
| Bond Fluctuation Model | Crosslinking is dependent on reaction kinetics, based on the number of | Fluctuations of monomers in chains is explicitly considered in the simulation | Individual monomer positions are assumed to be on an | The simulation explicitly considers excluded volume | Coarse-graining is on the level of a monomer and crosslinker. Hence, all monomers in | Crosslinking is based on the spatial position of unreacted chain ends | The starting monomers of each chain is placed stochastically in the |

| available neighbors around each unreacted chain and crosslinker | method, thus accounting for diffusion effects | underlying lattice | effects by requiring that no two monomers can occupy the same lattice site | every chain is tracked, thus making this simulation method computationally more expensive than all the other ones described above | around crosslinkers | simulation box, and then chain growth steps are run to form the polymer chains. However, since fluctuations of monomers in space is explicitly considered, this initial bias is eliminated to a large extent |
|---|---|---|---|---|---|---|

## VIII. Calculation of lattice length scale for the DWS networks:

Number of divisions along each edge $= a$

Number of unit cells $= (a+1)^3$

Number of nodes in a diamond lattice $= 8 * (a+1)^3$

Number of chains $= 16 * (a+1)^3$

Volume of simulation box $V = \frac{number\ of\ chains}{concentration} = \frac{16*(a+1)^3}{c} = \frac{16*(a+1)^3}{\frac{cR^3}{(Nb^2)^{1.5}}}$

Length of simulation box $L = V^{\frac{1}{3}} = b\sqrt{N}(a+1)\left[\frac{16}{cR^3}\right]^{\frac{1}{3}}$

Contour length $L_c = bN$

Lattice parameter $a_L = \left[\frac{L}{a+1}\right] = \frac{1}{\sqrt{N}}\left[\frac{16}{cR^3}\right]^{\frac{1}{3}}$

First nearest neighbor distance $d_{1,nn} = 0.433 * a_L = \frac{0.433}{\sqrt{N}} \left[\frac{16}{cR^3}\right]^{\frac{1}{3}}$

Second nearest neighbor distance $d_{2,nn} = 0.2387 * a_L = \frac{0.2387}{\sqrt{N}} \left[\frac{16}{cR^3}\right]^{\frac{1}{3}}$

For $N = 12$ and $cR^3 = 10$:

Lattice parameter $a_L = 0.337 L_c$

First nearest neighbor distance $d_{1,nn} = 0.146 L_c$

Second nearest neighbor distance $d_{2,nn} = 0.239 L_c$

## IX. Code validations for the loop counting algorithm

**Table S3:** Ideal networks used as test cases to validate the loop counting algorithm

| **Network** | **Vertex symbol** |
| --- | --- |
| Cayley tree | None (since there are no loops in such a structure) |
| Diamond lattice | $6_2.6_2.6_2.6_2.6_2.6_2$ |
| Graphite network | $6_1.6_1.6_1$ |
| Fullerene network | $5_1.6_1.6_1$ |

**Table S4:** Examples of micronetworks used as test cases for the loop counting algorithm

| **Network** | **Node under consideration** | **Vertex symbol** |
| --- | --- | --- |
|  |  |  |
| 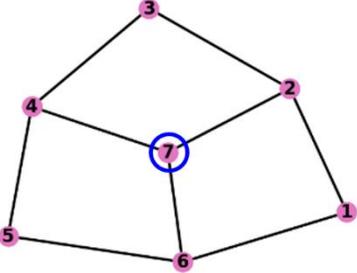 | 7 | 4.4.4 |
| 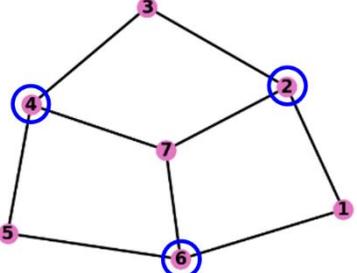 | 2,4,6 | 4.4.6 |
| 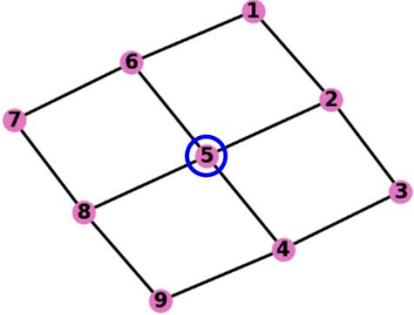 | 5 | 4.4.4.4.*.* |
| 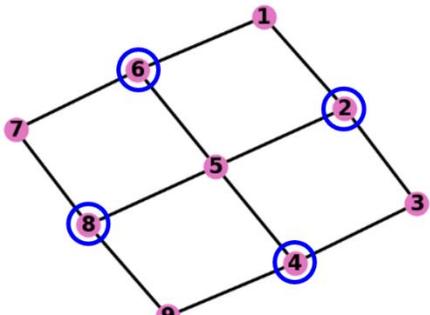 | 2,4,6,8 | 4.4.* |
| 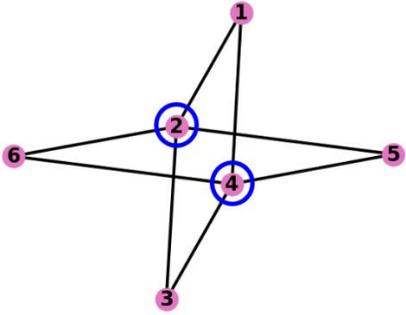 | 2,4 | 4.4.4.*.*.* |
| 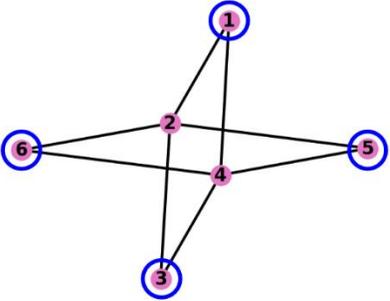 | 1,3,5,6 | $4_3$ |
| 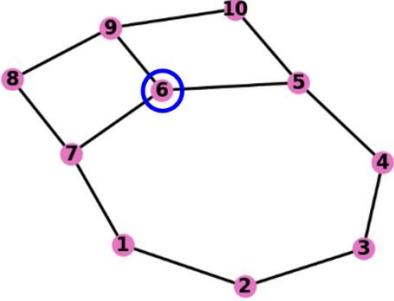 | 6 | 4.4.7 |
| 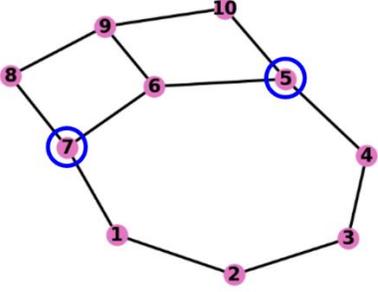 | 5,7 | 4.7.9 |
| 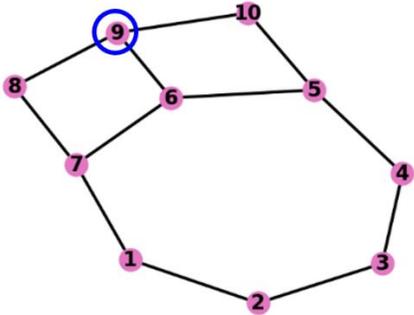 | 9 | 4.4.* |
| 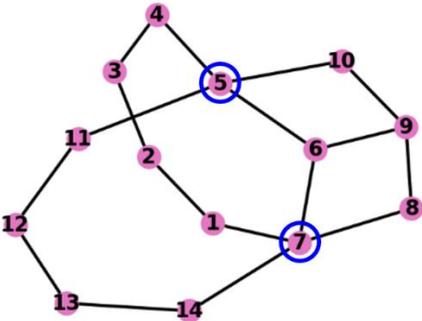 | 5,7 | 4.7.7.9.*.* |
| 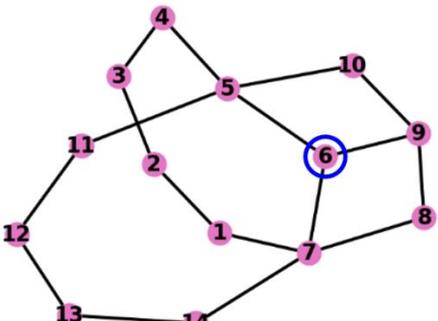 | 6 | 4.4.7$_2$ |
| 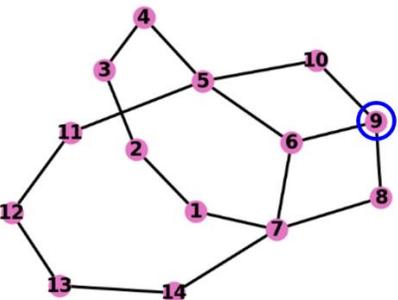 | 9 | 4.4.* |
| 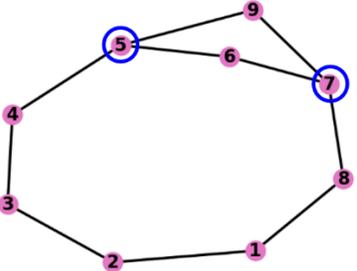 | 5,7 | 4,8.* |
| 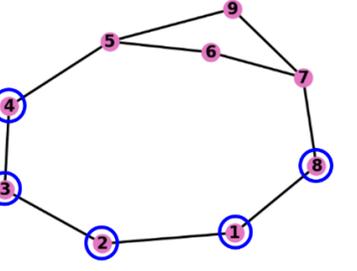 | 1,2,3,4,8 | $8_2$ |
| 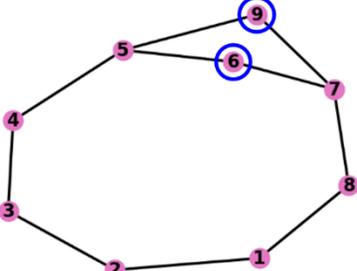 | 6,9 | $4_1$ |

## X. Hierarchical clustering analysis to identify network classes

A hierarchical clustering algorithm [11,12] is employed to identify three distinct network categories by assessing clusters in the pairwise EMD matrix. The algorithm initiates with a set of unused clusters in the formation of the hierarchy. When two clusters $s$ and $t$ are merged into a single cluster $u$, the clusters $s$ and $t$ are removed from the set and $u$ is added to it. This algorithm continues until only one cluster remains in the set, and this cluster becomes the root. Throughout this process, a distance matrix containing the distances between clusters is maintained, and is updated at each iteration to reflect the distance of the newly formed cluster $u$ with the remaining clusters in the set.

Ward's minimum variance criterion is used as a linkage method, and the distance between clusters $u$ and $v$, $d(u,v)$ is calculated as:

$$d(u,v) = \sqrt{\frac{|v|+|s|}{T}d(v,s)^2 + \frac{|v|+|t|}{T}d(v,t)^2 - \frac{|v|}{T}d(s,t)^2}$$

Where $u$ is the newly joined cluster consisting of clusters $s$ and $t$, $v$ is an unused cluster in the forest, $T = |v| + |s| + |t|$, and $|*|$ is the cardinality of its argument. The python library `scipy.cluster.hierarchy` has been used for this analysis.

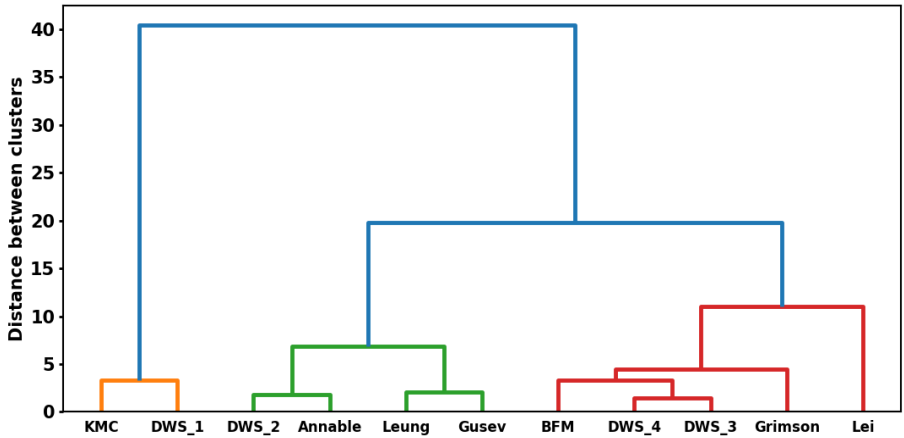

**Figure S9:** Dendrogram of the hierarchical clustering analysis of the network simulation algorithms considered in this work, performed using Ward's minimum variance criterion

**Code Availability:**

The computational codes for this work are available on the following Github repository:

https://github.com/olsenlabmit/Cyclic_topology_using_3D_Nets